\documentclass[runningheads]{llncs}
\usepackage{graphicx}
\usepackage{amsfonts}
\usepackage{amsmath}
\usepackage{booktabs}
\usepackage{multirow}

\usepackage[colorlinks, citecolor=blue]{hyperref}

\begin{document}
%
\title{FusionINN: Decomposable Image Fusion for Brain Tumor Monitoring}

\author{Nishant Kumar\inst{1}* \and Ziyan Tao\inst{1} \and Jaikirat Singh\inst{1} \and Yang Li\inst{2} \and Peiwen Sun\inst{3} \and Binghui Zhao\inst{3} \and Stefan Gumhold\inst{1}}
%

\institute{$^{1}$Chair of Computer Graphics and Visualization, Faculty of Computer Science, Technische Universität Dresden, Dresden, Germany \\ 
$^{2}$School of Computer Science and Engineering, Shandong University of Science and Technology, Qingdao, China \\
$^{3}$Department of Radiology, Shanghai Tenth People's Hospital, Tongji University Medical School, Shanghai, China \\
\email{$^{*}$nishant.kumar@tu-dresden.de}}
\maketitle              
\begin{abstract}
Image fusion typically employs non-invertible neural networks to merge multiple source images into a single fused image.~However, for clinical experts, solely relying on fused images may be insufficient for making diagnostic decisions, as the fusion mechanism blends features from source images, thereby making it difficult to interpret the underlying tumor pathology.~We introduce FusionINN, a novel decomposable image fusion framework, capable of efficiently generating fused images and also decomposing them back to the source images.
~FusionINN is designed to be bijective by including a latent image alongside the fused image, while ensuring minimal transfer of information from the source images to the latent representation.~To the best of our knowledge, we are the first to investigate the decomposability of fused images, which is particularly crucial for life-sensitive applications such as medical image fusion compared to other tasks like multi-focus or multi-exposure image fusion.~Our extensive experimentation validates FusionINN over existing discriminative and generative fusion methods, both subjectively and objectively.~Moreover, compared to a recent denoising diffusion-based fusion model, our approach offers faster and qualitatively better fusion results.~The source code of the FusionINN framework is available at:~\url{https://github.com/nish03/FusionINN}.


\end{abstract}
\section{Introduction}
\begin{figure}[!t]
	\centering
	\includegraphics[width=\linewidth]{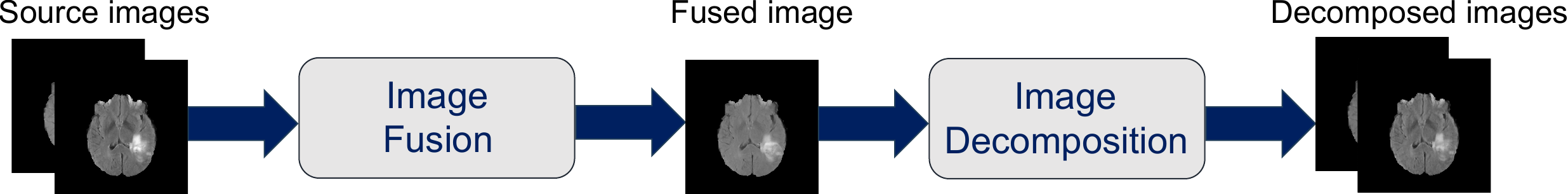}
	\caption{An illustration of the task of image fusion and decomposition.}
	\label{fig:task}
\end{figure}

Magnetic Resonance Imaging (MRI) techniques, such as Diffusion-weighted imaging with Apparent Diffusion Coefficient (DWI-ADC) and T2-weighted Fluid Attenuated Inversion Recovery (T2-Flair), offer invaluable insights into the intricate pathology of tumors.~A high-intensity signal on the T2-Flair image provides anatomical information about the presence of tumor and its boundary~\cite{Bitar}.~In contrast, DWI-ADC assists in revealing the tumor category, as a high-intensity signal indicates the existence of liquid components, i.e., necrotic tumor tissues and a low-intensity signal suggests the presence of solid components, i.e., enhancing tumor tissues~\cite{Sertoli}.~Clinicians commonly utilize such image modalities post-operatively to detect any residual necrotic tumor tissues and assess the potential for its recurrence by locating enhancing tumor tissues.~Fused images can aid in the visualization of the clinical features from multiple sources.~However, merging grayscale values can obscure salient features, thereby complicating clinical interpretation of the fused image.~To address this problem, we introduce the extended fusion task illustrated in Fig.~\ref{fig:task}, which demands decomposability of the fused image into the source images.


Prior works in image fusion leverage deep learning algorithms via discriminative training~\cite{deepfuse,masknet,funfusean,fusionCNN,fmivis,IFCNN,dlfusion,kumar2019multimodal,fusevis} or generative modeling using generative adversarial networks (GANs)~\cite{fusionGAN}.~However, the network architecture of such image fusion approaches is not invertible.~As a result, they have not been utilized for decomposing fused images.~Recently, a pre-trained Denoising Diffusion based image fusion model~\cite{DDFM}  has been proposed, that conditions each of the denoising diffusion steps on source images.~In principle, diffusion models allow stable training dynamics, while not suffering from mode collapse.~However, the decomposability of the fused images is also not explored in~\cite{DDFM}, possibly because the pre-trained UNet~\cite{unet} model used to perform the denoising steps is not invertible.~Additionally, diffusion models perform slow sequential sampling through multiple denoising steps to obtain the fusion output, due to which a real-time inference scheme is impractical.


We present normalizing flows as the generative model for medical image fusion and capitalize on their inherent invertibility to facilitate the decomposability of the fusion process.~The flow demonstrates efficient sampling capabilities and stability during training through the use of invertible transformations, which are beneficial for computer vision tasks~\cite{ffs,quantod}.~Previous attempts utilizing invertible neural networks (INNs) for image fusion~\cite{INN2,INN3,INN1,INN4,INN5} have predominantly integrated INNs only as a sub-module within a multi-step pipeline, preventing the invertibility of the end-to-end fusion procedure.~Notably, no prior studies have explored solving both the tasks of image fusion and decomposition through an end-to-end INN model.~The primary contributions of this work are as follows:
\begin{itemize}
    \item We introduce a first-of-its-kind image fusion framework, FusionINN, that harnesses invertible normalizing flow for bidirectional training.~
    FusionINN not only generates a fused image but can also decompose it into constituent source images, thus enhancing the interpretability for clinical practitioners.
    \item We present an extensive evaluation study that shows state-of-the-art results of FusionINN with common fusion metrics, alongside its additional capability to decompose the fused images.
    \item We also illustrate the effectiveness of FusionINN in fusing and decomposing images from clinical modalities that were not encountered during training.
\end{itemize}

\begin{figure}[!t]
	\centering
	\includegraphics[width=\linewidth]{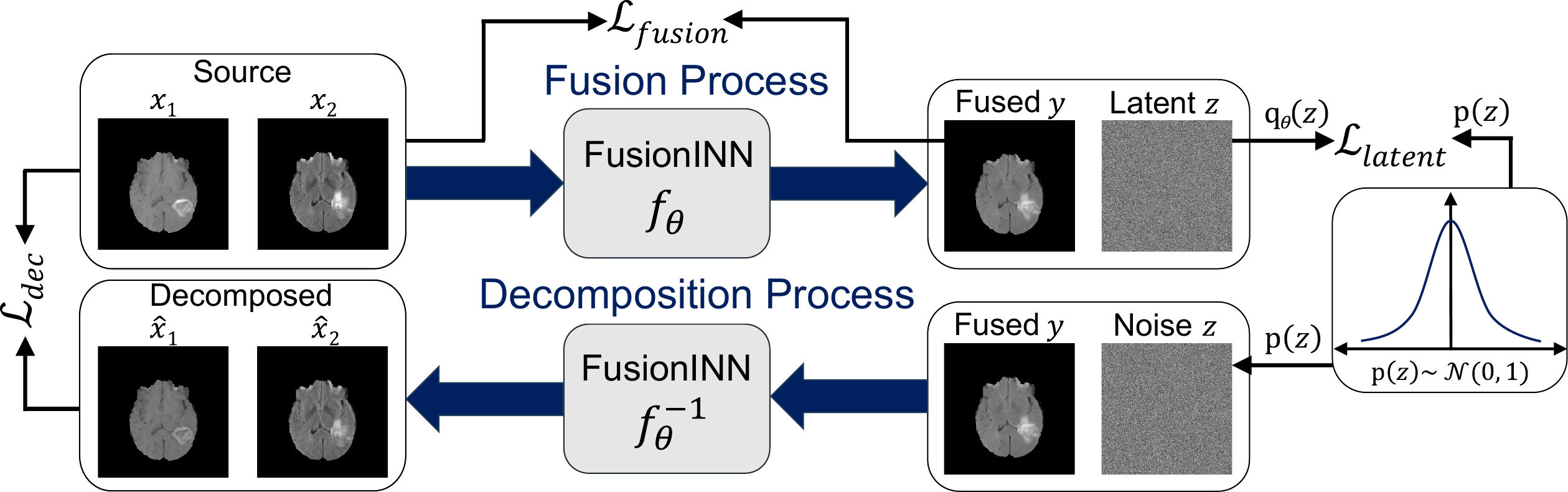}
	\caption{An overview of the FusionINN framework.}
	\label{fig:fusionINN}
\end{figure}

\section{Method}

The objective under decomposable image fusion, as depicted in Fig.~\ref{fig:task}, is to generate a fused image that closely resembles the source images and can be decomposed back into those source images without additional information.

\subsection{INN-based Decomposable Image Fusion}

The FusionINN framework for decomposable image fusion is shown in Fig.~\ref{fig:fusionINN}.~In the forward fusion process, the FusionINN transforms the two source images $x_1 \in \mathbb{R}^n$ and $x_2 \in \mathbb{R}^n$ to a fused image $y \in \mathbb{R}^n$ and a latent image $z \in \mathbb{R}^n$ using the normalizing flow network $f$ with parameters $\theta$ such that $[y,z] = f_{\theta}(x_1,x_2)$, where $n$ is the number of pixels in the four equal resolution images.~Consequently, the dimensionality of $[y,z]$ matches $[x_1,x_2]$ with $f^{}_\theta, f^{-1}_\theta:\mathbb{R}^{2n} \leftrightarrow \mathbb{R}^{2n}$.~Unlike GANs, which adversarially train two separate neural networks, normalizing flow requires training only a single network.~This simplifies the training  and makes it more stable, as there is no adversarial training dynamics.~We introduce the latent image $z$ to ensure the decomposability of the fused image, as the reverse mapping from a fused image to two source images is ill-posed.~As the latent image is unknown for the decomposition task, we aim to capture as few source image features as possible in it.~Therefore, we define the latent image $z$ to follow a multivariate normal distribution, such that $z \sim p(z) = \mathcal{N}(z; 0, I)$.~However, other design choices, such as a constant image $z$, are also feasible.~Finally, the decomposition process utilizes a newly sampled latent image $z$ along with the fused image $y$ through the reverse direction of FusionINN i.e., $f_{\theta}^{-1}$ to produce the decomposed images $\hat{x}_1$ and $\hat{x}_2$, such that $[\hat{x}_1,\hat{x}_2] = f^{-1}_\theta([y,z])$.~The inverse function $f_{\theta}^{-1}$ should learn to decompose the fused image $y$, independently from the latent image $z$, while ensuring that the decomposed images $\hat{x}_1$ and $\hat{x}_2$ closely resemble the source images $x_1$ and $x_2$.

\subsection{INN Architecture}

The FusionINN as a normalizing flow network 
$f_\theta$  consists of $k$ invertible coupling blocks stacked together such that $f = f_k \circ ... f_j \circ ... f_1$ with $[\hat{x}_1,\hat{x}_2] = f_\theta^{-1}(y,z)$ and $[y,z] = f_\theta(x_1,x_2)$.~In~\cite{dinh2017density}, the
coupling blocks consist of learnable affine functions, namely scaling ($s_1$ and $s_2$) and translation ($t_1$ and $t_2$).~We define these functions as convolutional neural networks (CNNs) with two convolutional layers, each followed by a ReLU activation.~The input to an arbitrary $j^{th}$ coupling block is first split into two parts $u^{j}_1$ and $u^{j}_2$, which are transformed by $s_1, t_1$ and $s_2, t_2$ networks that share the learnable parameters.~The output of the $j^{th}$ coupling block is the concatenation of the resulting parts $v^{j}_1$ and $v^{j}_2$ given as:

\begin{equation}
     v^{j}_1 = u^{j}_1 \odot \exp \big(s_2(u^{j}_2)\big) + t_2(u^{j}_2), \ \ \ \\ 
    v^{j}_2 = u^{j}_2 \odot \exp \big(s_1(v^{j}_1)\big) + t_1(v^{j}_1) 
    \label{eq1}
\end{equation}
 
where $\odot$ is the element-wise multiplication, and the exponential term ensures non-zero coefficients.~By construction, such a transformation is invertible, and $u^{j}_1, u^{j}_2$ can be recovered from $v^{j}_1,v^{j}_2$ (see~\cite{INN}).
~Between each coupling block, we implement a random permutation operation to reorganize the two channels obtained from the output of the previous block.~This permutation is applied only once and remains fixed during the training of FusionINN's learnable parameters \(\theta\).~Furthermore, following the channel permutation, we utilize an invertible downsampling operator~\cite{i-revnet} to reduce the spatial resolution of the input channels without losing any information.~For example, when $k = 3$, an invertible downsampling operation precedes the second coupling block, and an invertible upsampling operation is applied before the third coupling block to maintain the resolution of the final output of the normalizing flow network $f_\theta$.~This operation enables the network to increase its receptive field and effectively capture features at multiple scales.~We also apply a sigmoid function as the final layer of the network to obtain the normalized fused image $y$.

\subsection{Unsupervised Learning}

The learning scheme of our FusionINN framework, depicted in  Fig.~\ref{fig:fusionINN}, operates without a predefined fusion groundtruth.~Therefore, we approach the fusion task as a fully unsupervised problem, utilizing the fusion loss $\mathcal{L}_{fusion}$. This loss function allows FusionINN to optimize the fused image without explicit supervision, learning directly from the source images.~Additionally, FusionINN learns to shape the latent image $z$ to conform to a standard normal distribution through the $\mathcal{L}_{latent}$ loss, which minimizes information transfer from source images to the latent image.~We also define a decomposition loss as $\mathcal{L}_{dec}$, which aids in estimating the source images from the fused image.~With these loss functions, our FusionINN framework not only achieves superior fusion results but also facilitates image decomposition.

\subsubsection{Fusion Loss:}~To learn the fused image $y$ from the flow network $f_\theta$ in an unsupervised manner, we follow~\cite{deepfuse} and leverage the metric Structural Similarity Index ($Q_{SSIM}$)~\cite{structuralsimilarity} as the differentiable loss function to maximize the similarity between the source and the fused images.~The loss function is formulated as:

\begin{equation}
    \begin{gathered}
         \mathcal{L}_{SSIM} = \{1 - Q_{SSIM}(x_1, y)\} +  \{1 - Q_{SSIM}(x_2, y)\}
    \end{gathered}
    \label{eq3}
\end{equation}


The sub-loss terms in $\mathcal{L}_{SSIM}$ are subtracted from 1 to satisfy the loss minimization objective, as $Q_{SSIM}$ computes the similarity between the two images.~However, while $Q_{SSIM}$ is effective in preserving the structure and the contrast of an image, it can alter the brightness and make the image appear duller, as discussed in~\cite{ssim_l2}.~To address this, we use the squared $\ell_2$ loss in addition to the $Q_{SSIM}$ metric to better preserve the luminance of the fused image, as squared $\ell_2$ loss directly penalizes differences in pixel intensities.~Finally, given the weightage parameter as $\lambda$, the $\mathcal{L}_{\ell_2}$ and $\mathcal{L}_{fusion}$ losses are expressed as:

\begin{equation}
\begin{gathered}
\mathcal{L}_{\ell_{2}} = ||y - x_1||^{2}_2 + ||y - x_2||^{2}_2, \ \ \ 
 \mathcal{L}_{fusion} = \{\lambda \mathcal{L}_{SSIM} + (1 - \lambda)  \mathcal{L}_{\ell_2}\}
 \end{gathered}
 \label{eq4}
\end{equation}


\subsubsection{Latent Loss:}
We model the distribution of the latent image $z$ with a multivariate Gaussian $p(z)= \mathcal{N}(z; 0, I)$.~We utilize Maximum Mean Discrepancy ($\mathrm{MMD}$)~\cite{MMD} as the loss function to quantify the difference between the probability distribution $p(z)$ and the distribution $q_\theta(z)$ of the latent image $z$ generated by the forward process of the FusionINN model, $f_\theta$.~Consequently, the latent loss $\mathcal{L}_{latent}$ is defined as 
$\mathcal{L}_{latent} =  \mathrm{MMD}(q_\theta(z) \mkern3mu \Vert \mkern3mu p(z))$.
~This enables the learned distribution $q_\theta(z)$ to be approximated as the standard normal distribution $p(z)$ after minimization of the $\mathcal{L}_{latent}$ loss.

\subsubsection{Decomposition Loss:}

~We define the decomposition loss $\mathcal{L}_{dec}$ in the reverse direction of $f_\theta$ i.e. $f^{-1}_\theta$ to decompose the fused image $y$ back to the source images, using a newly sampled latent image $z$.~We implement the $\mathcal{L}_{dec}$ loss as the combination of the $\mathcal{L}^{SSIM}_{dec}$ and $\mathcal{L}^{\ell_2}_{dec}$ losses, which are weighted using the meta-parameter $\lambda$, similar to the $\mathcal{L}_{fusion}$ loss.~The $\mathcal{L}^{SSIM}_{dec}$ loss employs the $Q_{SSIM}$ metric, while $\mathcal{L}^{\ell_2}_{dec}$ computes the squared $\ell_2$-loss to measure the dissimilarity between the decomposed and source images.~Hence, given the decomposed images $\hat{x}_1$ and $\hat{x}_2$, the losses $\mathcal{L}^{SSIM}_{dec}$, $\mathcal{L}^{\ell_2}_{dec}$ and $\mathcal{L}_{dec}$ are computed as:

\begin{equation}
    \begin{gathered}
         \mathcal{L}^{SSIM}_{dec} =  \{1 - Q_{SSIM}(x_1, \hat{x}_1)\} + \{1 - Q_{SSIM}(x_2, \hat{x}_2)\} \\
         \mathcal{L}^{\ell_{2}}_{dec} = ||\hat{x}_1 - x_1||^{2}_2 + ||\hat{x}_2 - x_2||^{2}_2,  \ \ \
         \mathcal{L}_{dec} = \lambda \mathcal{L}^{SSIM}_{dec} + (1 - \lambda) \mathcal{L}^{\ell_2}_{dec}
    \end{gathered}
    \label{eq6}
\end{equation}

\begin{figure*}[!t]
	\centering	\includegraphics[width=\linewidth]{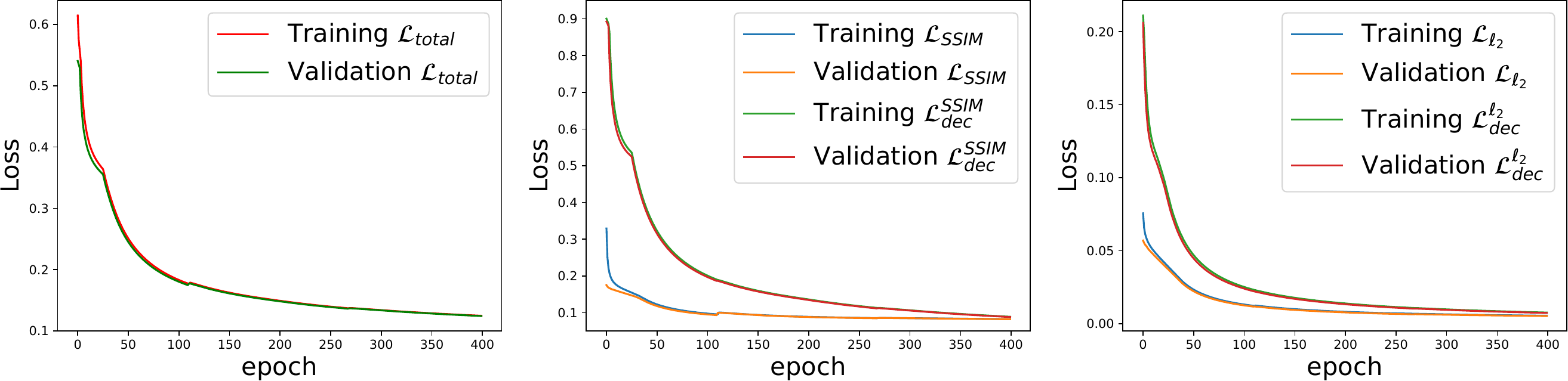}
	\caption{Loss curves for a FusionINN training instance with $k = 4$, $\lambda = 0.9$ and $\alpha = 0.5$.}
	\label{fig:loss}
\end{figure*}%

\subsubsection{Total Loss:}In the forward process, the FusionINN optimizes the mapping $[y, z] = f_{\theta}(x_1,x_2)$ using $\mathcal{L}_{fusion}$ and $\mathcal{L}_{latent}$ losses.~Additionally, FusionINN's invertibility guarantees that the latent image generated from the forward  process can precisely reproduce the source images.~However, during the reverse process, we sample a new latent image $z$ from the normal distribution $p(z)$ to maximize the decomposition accuracy, independent from any specific choice of $z$.~The resampled latent image $z$, together with the fused image $y$ is used to perform the reverse process by optimizing the mapping $[\hat{x}_1,\hat{x}_2] = f^{-1}_{\theta}(y, z)$ via the $\mathcal{L}_{dec}$ loss.~Finally, we weight the forward losses i.e., $\mathcal{L}_{fusion}$ and $\mathcal{L}_{latent}$ as well as the decomposition loss i.e., $\mathcal{L}_{dec}$ using the parameter $\alpha$ and formulate the total loss function $\mathcal{L}_{total}$ as follows: 

\begin{equation}
    \begin{gathered}
         \mathcal{L}_{total} =   \{\alpha (\mathcal{L}_{fusion} + \mathcal{L}_{latent}) + (1- \alpha) \mathcal{L}_{dec}\}  \\
    \end{gathered}
    \label{eq7}
\end{equation}

\subsubsection{Training Procedure:} 

We learn the FusionINN's parameters $\theta$ by iteratively optimizing them to minimize the total loss function, $\mathcal{L}_{total}$.~This involves computing the gradients of \(\mathcal{L}_{total}\) with respect to each parameter using backpropagation and updating the parameters using Adam optimization~\cite{adam} with a learning rate of \(3 \times 10^{-4}\).~The training is performed over $400$ epochs with a batch size of $64$. We also utilize a learning rate scheduler to reduce the learning rate by a factor of $0.95$ if the validation loss does not improve for eight epochs, preventing the model from getting stuck in local minima and ensuring smoother convergence.~The loss curves for \(\mathcal{L}_{total}\) and the sub-losses \(\mathcal{L}_{SSIM}\), \(\mathcal{L}^{SSIM}_{dec}\), $\mathcal{L}_{\ell_2}$ and $\mathcal{L}^{\ell_2}_{dec}$ at each training epoch are illustrated in Fig.~\ref{fig:loss}.

\section{Results and Discussion}
\subsubsection{Data Description:}~We use the publicly available BraTS-2018 brain imaging dataset~\cite{brats} to prepare our training and validation data.~We extract post-contrast T1-weighted (T1-Gd) and T2-Flair as the two source images, acquired with different clinical protocols and different scanners from multiple medical institutions.~The data has been pre-processed, i.e., co-registered to the same anatomical template, interpolated to the same resolution and skull-stripped~\cite{brats}.~We only extract those images from the dataset where the clinical annotation comprises of the necrotic core, non-enhancing tumor, and the peritumoral edema.~This results in 9653 image pairs of T1-Gd and T2-Flair modalities.~We randomly assign 8500 image pairs as training and 1153 image pairs as the validation set.

\begin{table}[!htb]
\centering
\caption{%
Comparison of the fusion performance of the evaluated models on the validation set of our pre-processed BraTS-2018 images~\cite{brats}.~The results from each model show averaged scores from five metrics after comparing the fused images with the source image pairs.~For each metric, the best-performing model is highlighted in bold.
}
\scalebox{0.95}{
\begin{tabular}{c c c c c c c}
\toprule
        Model Type & Model Name & \multirow{1}*{$Q_{SSIM}$ $\uparrow$} & \multirow{1}*{$Q_{FMI}$ $\uparrow$} & \multirow{1}*{$Q_{NCIE}$ $\uparrow$} & \multirow{1}*{$Q_{XY}$ $\uparrow$}  & \multirow{1}*{$Q_{P}$  $\uparrow$}\\
	\toprule
	Discriminative  & DeepFuse~\cite{deepfuse} & $0.927$ & $0.791$ & $0.806$  & $0.449$ & $0.766$ \\
	(Equal Dimension)	 & FunFuseAn~\cite{funfusean} & $0.930$ & $0.845$ & $0.806$ & $0.481$ & $0.781$ \\
	\midrule
	Discriminative  & Half-UNet~\cite{half_unet} & $0.933$ & $0.850$ & $0.805$ & $0.464$ & $0.774$ \\
	(Dimension Reduction)  & UNet~\cite{unet} & $0.934$ & $0.835$ & $0.805$ & $0.420$ & $0.711$ \\
	& UNet++~\cite{unet++} & $\mathbf{0.937}$ & $0.849$ & $0.805$ & $0.433$ & $0.739$ \\
	 & UNet3+~\cite{unet3+} & $0.937$ & $0.849$ & $0.805$ & $0.434$ & $0.742$ \\
    \midrule
	Generative & DDFM~\cite{DDFM} & $0.921$ & $\mathbf{0.861}$ & $0.806$ & $0.472$ & $0.702$ \\
                   & \textbf{FusionINN~(Ours)} & $0.927$ & $0.835$ & $\mathbf{0.806}$ & $\mathbf{0.493}$ & $\mathbf{0.783}$ \\
	\bottomrule
\end{tabular}
}
\label{table:main}
\end{table}%
\begin{figure*}[!htb]
	\centering	\includegraphics[width=\linewidth]{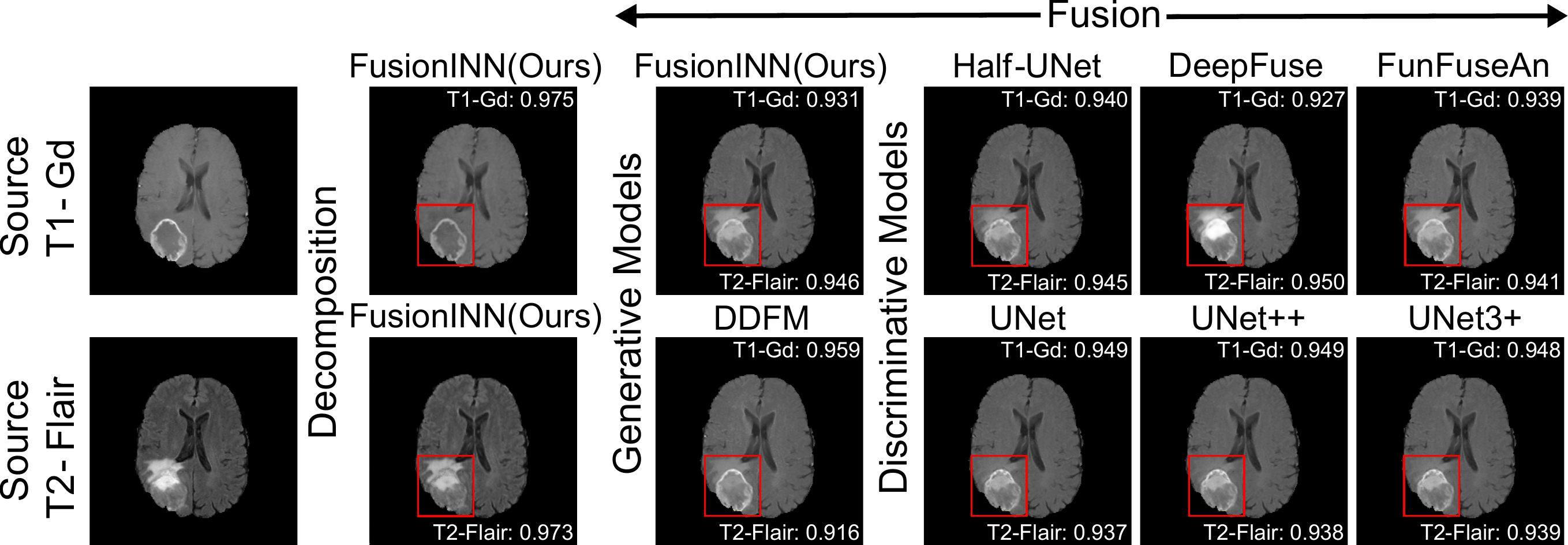}
	\caption{Fusion results obtained from the evaluated models on a sample validation image pair.~The $Q_{SSIM}$ scores for individual modalities are shown in the fused images.}
	\label{fig:fusion}
\end{figure*}%

\subsubsection{Competitive Methods and Evaluation Metrics:}We assess FusionINN's performance by comparing it with other fusion methods, namely DeepFuse~\cite{deepfuse} 
and FunFuseAn~\cite{funfusean}.~We also repurpose popular image segmentation models namely Half-UNet~\cite{half_unet}, UNet~\cite{unet}, UNet++~\cite{unet++}, and UNet3+\cite{unet3+} for the image fusion task.~Each of these models involve discriminative modeling and are trained on the fusion loss, i.e., $\mathcal{L}_{fusion}$ (Eq.~\ref{eq4}).~These models are non-invertible and can only be used to estimate fused images.~We maintain a common benchmark of meta-parameters during training of these models.~Furthermore, we employ the pre-trained Denoising Diffusion-based Fusion model (DDFM)~\cite{DDFM} as a generative method to evaluate its performance on our validation images.~We utilize five quantitative metrics specifically designed for assessing the image fusion quality.~The metrics are Feature Mutual Information ($Q_{FMI}$)~\cite{fmi}, Structural Similarity Index ($Q_{SSIM}$)~\cite{structuralsimilarity}, Non-linear Correlation Information Entropy ($Q_{NCIE}$)~\cite{ncie}, and by Petrovic et al. ($Q_{XY}$)~\cite{xydeas}, and Piella et al. ($Q_{P}$)~\cite{piella}.~The metric $Q_{XY}$ use gradient representation of the source images to quantify the information or feature transfer to the fused images.~On the other hand, $Q_{P}$ weights the structural similarity scores based on local saliencies of the two source images.

\subsubsection{Fusion and Decomposition Performance:} Table~\ref{table:main} presents the quantitative fusion results of the evaluated models across various fusion metrics after averaging over the validation images.~Our FusionINN model demonstrates either comparable or superior fusion performance with respect to all other evaluated models across metrics such as $Q_{NCIE}$, $Q_{XY}$, and $Q_P$.~Notably, FusionINN also exhibits competitive results on $Q_{SSIM}$ metric.~The Fig.~\ref{fig:fusion} shows qualitative fusion results using a sample image pair from the validation set.~The fusion results from the FusionINN model is competitive with other methods, and its decomposition results closely resemble the source images.
~Despite UNet-based methods exhibiting comparable $Q_{SSIM}$ scores, FusionINN excels in preserving the high-intensity features from the T2-Flair image within the fused output. 

\begin{table}[!htb]
\centering
\caption{The effect of coupling blocks $k$, latent image $z$, and parameters $\lambda$ and $\alpha$ on the fusion and decomposition performance is examined. The results are obtained from a single training run, using different initial random seeds for each combination of meta-parameters. These results are averaged $Q_{SSIM}$ scores over the validation images, with $Q_{SSIM}(x, y)$ for fusion and $Q_{SSIM}(x, \hat{x})$ for decomposition. When studying $\alpha$, $\lambda$, and $k$, we maintain $z \sim \mathcal{N}(0, I)$. Additionally, we fix $k = 3$ when analyzing the impact of different types of latent image $z$ on the fusion and decomposition results.}
\scalebox{0.8}{
\begin{tabular}{c c c c c}
\toprule
        Weight~($\alpha$) & \multicolumn{2}{c}{Fusion} & \multicolumn{2}{c}{Decomposition}\\
        ($k = 3, \lambda = 0.8$) & T1-Gd & T2-Flair & T1-Gd & T2-Flair \\
	\toprule
	0.2  & 0.903 & 0.929 & 0.930 & 0.930 \\
	0.5  & 0.921 & \textbf{0.933} & \textbf{0.976} & \textbf{0.972} \\
    0.8  & 0.926 & 0.933 & 0.927 & 0.920 \\
    1.0  & \textbf{0.948} & 0.898 & 0.033 & 0.004 \\
	\bottomrule
\end{tabular}
}
\scalebox{0.8}{
\begin{tabular}{c c c c c}
\toprule
        Weight~($\lambda$)& \multicolumn{2}{c}{Fusion} & \multicolumn{2}{c}{Decomposition}\\
        ($k = 3, \alpha = 0.5$) & T1-Gd & T2-Flair & T1-Gd & T2-Flair \\
	\toprule
	0.8    & 0.921 & \textbf{0.933} & \textbf{0.976} & 0.972 \\
	0.9    & 0.925 & 0.929 & 0.974 & \textbf{0.974} \\
    0.99   & 0.915 & 0.928 & 0.969 & 0.974 \\
    0.999  & \textbf{0.935} & 0.923 & 0.937 & 0.920 \\
	\bottomrule
\end{tabular}
}
\scalebox{0.8}{
\begin{tabular}{c c c c c}
        Blocks~($k$) & \multicolumn{2}{c}{Fusion} & \multicolumn{2}{c}{Decomposition}\\
        ($\alpha = 0.5, \lambda = 0.8$) & T1-Gd & T2-Flair & T1-Gd & T2-Flair \\
	\toprule
	  1  & 0.914 & \textbf{0.943} & 0.151 & 0.093 \\
	  2  & \textbf{0.935} & 0.923 & 0.945 & 0.928 \\
        3  & 0.921 & 0.933 & \textbf{0.976} & \textbf{0.972} \\
        4  & 0.923 & 0.936 & 0.937 & 0.939 \\
	\bottomrule
\end{tabular}
}%
\scalebox{0.8}{
\begin{tabular}{c c c c c}
        Latent~($z$) & \multicolumn{2}{c}{Fusion} & \multicolumn{2}{c}{Decomposition}\\
        ($\alpha = 0.5, \lambda =0.8$) & T1-Gd & T2-Flair & T1-Gd & T2-Flair \\
	\toprule
	0  & 0.918 & 0.930 & 0.932 & 0.928 \\
	$\mathcal{N}(0, I)$  & 0.921 & \textbf{0.933} & \textbf{0.976} & \textbf{0.972} \\
    $\mathcal{U}[0, I)$  & \textbf{0.924} & 0.925 & 0.958 & 0.954 \\
    1 & 0.916 & 0.929 & 0.967 & 0.969 \\
	\bottomrule
\end{tabular}
}
\label{table:parameter}
\end{table}
\begin{figure}[!htb]
	\centering	\includegraphics[width=\linewidth]{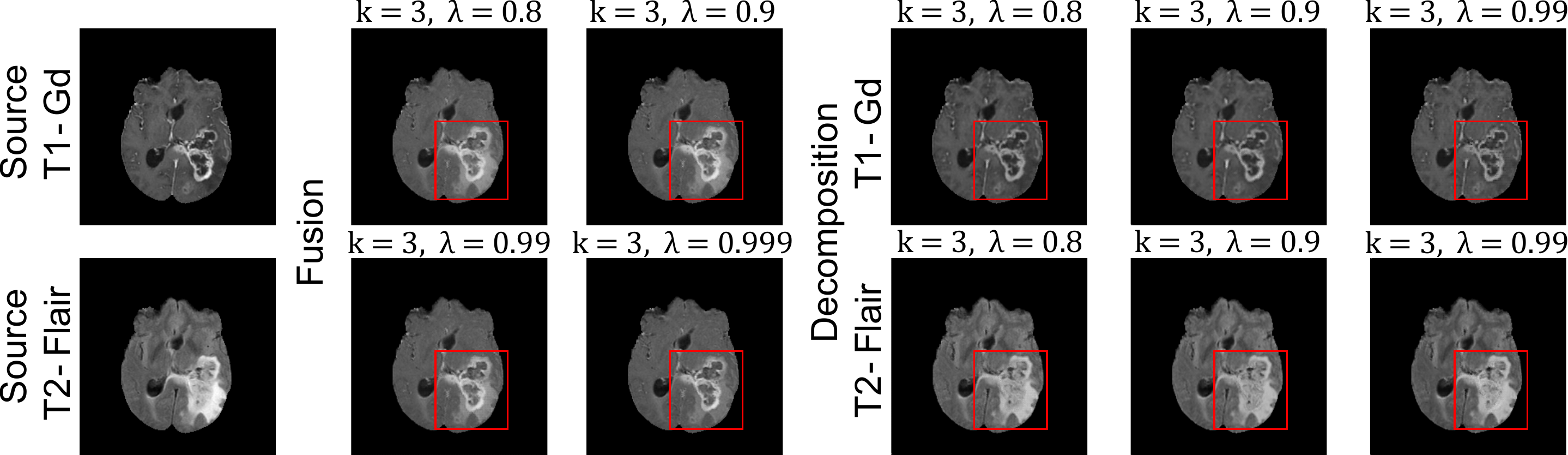}
	\caption{FusionINN results at $\alpha = 0.5$, $z \sim \mathcal{N}(0, I)$ and $k$ as number of coupling blocks.}
\label{fig:parameter_study}
\end{figure}%

\subsubsection{Ablation Studies:}

Table~\ref{table:parameter} demonstrates the impact of various parameters on FusionINN's fusion and decomposition performance.~The results in the upper left portion of Table~\ref{table:parameter} indicate that three coupling blocks with $\lambda = 0.8$ and $\alpha = 0.5$ produce competitive results in terms of $Q_{SSIM}$ scores.~Furthermore, increasing $\alpha$ enhances image fusion performance with respect to at least one source modality.~This can be attributed to a higher weightage given to the $\mathcal{L}_{fusion}$ loss in the optimization process.~We also explored different latent priors for $z$, including learning zeros, ones, and a uniform distribution $\mathcal{U}[0,1)$.~The results in the bottom right portion of Table~\ref{table:parameter} show that, on average, the fusion performance is similar under each type of latent prior for $z$.~This indicates that the latent image $z$ does not influence the construction and quality of the fused images.~Furthermore, interpreting the decomposition performance, it can be argued that, on average, a constant image $z$ with only ones in its pixel values performs almost as good as a latent image $z$ defined with random noise. In Fig.~\ref{fig:parameter_study}, the qualitative fusion and decomposition results portray that both $\lambda = 0.8$ and $0.9$ provide a good compensation of $Q_{SSIM}$ via squared $\ell_2$-loss, resulting in superior visual quality of the images.


\subsubsection{Clinical Translation:} 
\begin{figure}[!t]
	\centering
\includegraphics[width=\linewidth]{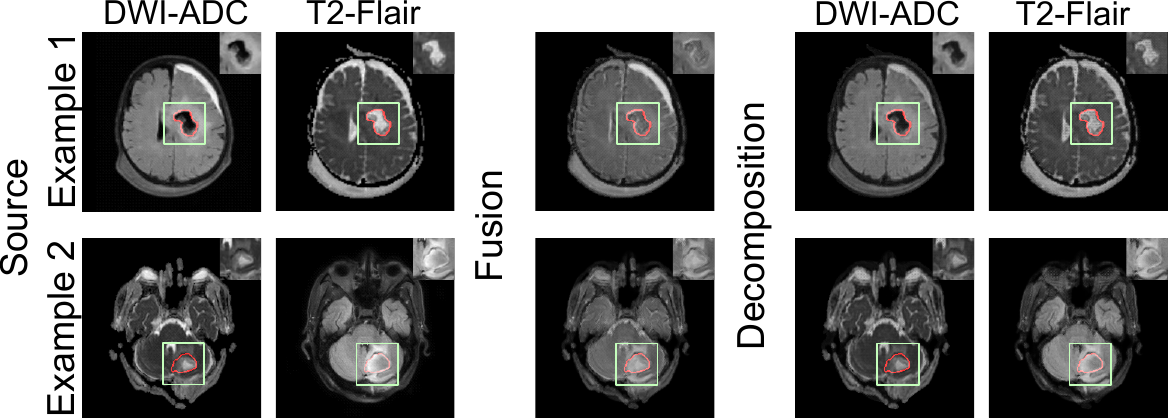}
	\caption{The results from the FusionINN model on  clinically acquired post-operative image pairs.~The clinicians annotated tumor boundaries (highlighted in red), and we display tumor features and their surroundings within green boxes on each image.}
\label{fig:clinical_study}
\end{figure}

In this study, we aimed to evaluate the robustness of the FusionINN model for practical clinical usage.~To achieve this, we assessed the model's performance on entirely new and clinically relevant test modalities that were not included in the training data.~Fig.~\ref{fig:clinical_study} illustrates clinically acquired image pairs from DWI-ADC and T2-Flair modalities, showing post-operative tumor regions of two patients following brain surgery.~The medical practitioners sought both fused and decomposed images of the test image pairs shown in Fig.~\ref{fig:clinical_study} to better evaluate the model's efficacy in aiding prognosis.~Specifically, the model was expected to produce images that clearly delineate features in T2-Flair indicative of the tumor's anatomical boundary, while also preserving high- and low-intensity DWI-ADC features related to residual necrotic and enhancing tumor tissues.~Note that the FusionINN model was trained on image pairs of T1-Gd and T2-Flair modalities.~The results shown in Fig.~\ref{fig:clinical_study} demonstrate that the model preserves salient features from both modalities in its decomposed images and effectively combines source features into the fused image.~These findings highlight the efficacy and generalization capability of the model to accurately construct fused and decomposed images, even for unseen test images from new image modalities.~Therefore, the clinically robust results obtained from the FusionINN model should assist clinicians in making better diagnostic decisions.


\section{Conclusion}
We introduced a novel framework that integrates the image decomposition task into the fusion problem through the utilization of an invertible and end-to-end normalizing flow network, thereby effectively addressing both optimization tasks with the same model.~The bidirectional trainability of FusionINN ensures the robust decomposition of fused images back to their source images using arbitrary latent image representations.~Our framework also showcases its capability in producing clinically relevant fusion and decomposition results.~Through extensive evaluation utilizing multiple image fusion metrics, FusionINN consistently achieves competitive results when compared to existing generative and discriminative models, while marking itself as the first framework to enable decomposability of fused images.~To promote reproducibility and further research, we encourage readers to access the FusionINN's source code via the link provided in the paper abstract.~Future work may involve learning the latent space not as random noise, but rather optimizing it for clinically useful tasks such as image segmentation.~Additionally, incorporating feedback from clinicians may help enhance the learning scheme for image fusion and decomposition to better align with specific clinical requirements.

\section*{Acknowledgments}
This work was primarily supported by the Center for Scalable Data Analytics and Artificial Intelligence (ScaDS.AI) Dresden/Leipzig, Germany.~The work was also partially funded by DFG as part of TRR~248 -- CPEC (grant 389792660) and the Cluster of Excellence CeTI (EXC2050/1, grant 390696704).~The authors gratefully acknowledge the Center for Information Services and HPC (ZIH) at TU Dresden for providing computing resources.

\end{document}